\documentclass[a4paper,fleqn,singleside,12pt]{article}
\usepackage{float}
\usepackage{graphicx}
\usepackage{amsmath} 
\usepackage{subfigure}
\usepackage{multirow}   
\usepackage{bm}
\usepackage{capt-of}      
\usepackage{amssymb,amssymb}   
\usepackage{rotating}
\usepackage[margin=1.0in]{geometry}
\usepackage{hyperref}
\usepackage{xcolor} 
\hypersetup{ colorlinks, citecolor=blue ,linkcolor=black }
\begin{document}
{\centering
\section*{Diffusion of Zwitterion Glycine, Diglycine, and Triglyine in Water}}
{\centering\textbf{Yadav Prasad Kandel and Narayan Prasad Adhikari${^*}$}\\}
{\centering{\it {Central Department of Physics, Tribhuvan University, Kathmandu, Nepal.}\\}}
{\centering {\it{${^*}$ email: npadhikari@tucdp.edu.np}\\}}
\begin{abstract}
\noindent Diffusion, transport of mass in response to concentration and thermal energy gradient, is an important transport property, vital in material science and life science. In the present work classical molecular dynamics study of diffusion of zwittterion glycine, zwitterion diglycine and zwitterion triglycine in water have been carried out. Self and binary diffusion coefficients of aqueous solution of these molecules have been calculated using Einstein method. Our results agree with experimental data reported in literatures. Temperature dependency of diffusion of glycine in water have been explored using estimated values of self and binary diffusion coefficients at four different temperatures. Effect of peptide bond formation in diffusion has been studied using peptide chain composed of up to three monomers of glycine. The structure of the system is analyzed using radial distribution function of different atom. 

\noindent Keywords:  Diffusion Coefficient, Molecular dynamics, RDF, Arrhenius behaviour, Glycine, Diglycine, Triglycine.
\end{abstract}
 
\section{Introduction}
\noindent Amino acids are the organic substances which contain both amino and acid groups. Out of 300 naturally found amino acids only 20 serve as building blocks of protein \cite{AminoDefn}. Among these 20 building blocks of protein, glycine (gly), the simplest of the amino acids, is an essential component of important biological molecules, a key substance in many metabolic reactions, the major inhibitory neurotransmitter in the spinal cord and brain stem, and an anti-inflammatory, cytoprotective, and immune modulating substance \cite{AAS:AAS786}. Glycylglycine or diglycine (dgl) and glycylglycylglycine or triglycine (tgl) are peptides of glycine haveing two and three monomer units in chain respectively \cite{Lehninger}.  Understanding of physical properties like diffusion, transfer of mass due to concentration and thermal gradient, can provide meaningful information about interatomic and intermolecular interactions. Knowledge of diffusion of amino acids also helps in the understanding the dynamics of protein and protein folding. To produce the amino acids in reaction and extraction/separation processes efficiently, it is of great importance to estimate the mass transfer rates and design optimum chemical reactors and separators \cite{Umecky}. Also, it is very crucial to understand the correlation between the structure of system and diffusion.\\

\noindent Study of diffusion of biomolecules in aqueous and other medium has attracted a significant number of researchers, both in experiment as well as in simulation. At 298.2 K temperature the binary diffusion coefficient of aqueous glycine, at infinitesimally dilute solution, has been reported by  Longsworth \cite{longsworth2} to be 10.55, Lyons and Thomas \cite{glydiff1} to be 10.64, Woolf et al. \cite{glydiff2} to be 10.59, and Ma et al. \cite{glydiff4} to be 10.62, all in the unit of $10^{-10} m^2.s^{-1}$. Umecky et al. \cite{Umecky} have studied the temperature dependency of binary diffusion coefficient of glycine and reported the diffusion coefficient to be 9.36 x $10^{-10} m^2.s^{-1}$at 293.2 K and increase in it by 12.83 x $10^{-10} m^2.s^{-1}$ corresponding to 40 K increase in temperature. Changwei et al. \cite{concenExp} have shown the binary diffusion coefficient of aqueous glycine depends upon concentration and changes from 10.4011 to 9.4258 x $10^{-10} m^2.s^{-1}$ corresponding change in concentration from 0.1057 to 0.9045 $mol.L^{-1}$. Campo \cite{campo} have carried molecular dynamics (MD) study of hydration and structure of glycine in water and other works have been done to study other different properties of glycine, diglycine, and triglycine. To our best knowledge, no realistic molecular dynamics simulation has been carried out, which mimics the experiment in best possible way, to estimate diffusion coefficient of aqueous solution of these molecules. In present work, self and binary diffusion coefficient of aqueous solution of zwitterion glycine has been estimated at different temperatures from molecular dynamics simulation, at the mole fraction scaled to match experimental values reported by Umecky et al. \cite{Umecky}, and Longsworth \cite{longsworth2}. Diffusion coefficients of aqueous solution of zwitterion diglycine and zwitterion triglycine have also been estimated. The structure of systems and its effect on diffusion have been discussed.\\

\noindent In molecular dynamics simulation, the experimental environment could be mimicked by modeling different interactions between the atoms and molecules \cite{sunilsir}. Therefore, it is considered to be one of the best alternative for study of system dynamics like diffusion and structure for it being economical and free from experimental hazards \cite{skthapa}.  In the recent years, the MD techniques have evolved in the application to macroscopic or real systems to study the complex dynamic processes that occur in biological systems like protein stability, protein folding and unfolding, conformational changes etc. \cite{DipendraPaper, mehrer}.

\noindent In this paper, we have discussed theoretical background of the present work in section two. Computational details are explained in section three. Results are presented and discussed in section four, and conclusions are presented in section five.

\section{Theory}
\noindent Diffusion is a spontaneous mass transport phenomena by which matter is carried from one part of the system to another as a result of random molecular motion \cite{ ippaper}. It takes place in an account of concentration inhomogeneity and thermal gradient. Diffusion is an essential function in living organisms and has great many application in modern material science and technology.\\

\noindent Diffusion of particle in homogeneous medium with no chemical concentration gradient is called the self diffusion coefficient. Rate of the self-diffusion is measured in terms of self-diffusion coefficient \cite{ksharma}. Under the assumption of isotropy of medium, if $r(t) - r(0)$ is the change of position of diffusing particle in time $t$, then the macroscopic transport property- self diffusion coefficient can be related to microscopic property-  mean squared displacement of material, by using Einstein's relation, as:
\begin{equation}
\centering
\label{selfdiff}
D = \lim_{t \to \infty}  \frac{\langle \lbrack r(t) - r(0) \rbrack ^2 \rangle}{6 t}
\end{equation}

\noindent Here, $\langle ... \rangle$ represents the ensemble average of quantity inside the angled bracket, which in our case is the square of displacement. Thus, the self diffusion coefficient of any species is one sixth of slope of mean squared displacement plotted as a function of time.\\

\noindent The diffusion of two different species in a binary mixture is called binary or mutual diffusion and the corresponding diffusion coefficient is called binary or mutual diffusion coefficient. If self diffusion coefficients of two individual species A and B is $D_A$ and $D_B$, with mole fraction $N_A$ and $N_B$ respectively, the binary diffusion coefficient $D_{AB}$ of these species, according to Darken's phenomenological relation, is \cite{darken}:
\begin{equation}
\centering
\label{darken}
D_{AB} = N_{B}D_{A} + N_{A}D_{B}
\end{equation}

\section{Computaitonal Details}
\begin{figure}[h]
\minipage{0.40\textwidth}
\centering
\includegraphics[scale=0.55]{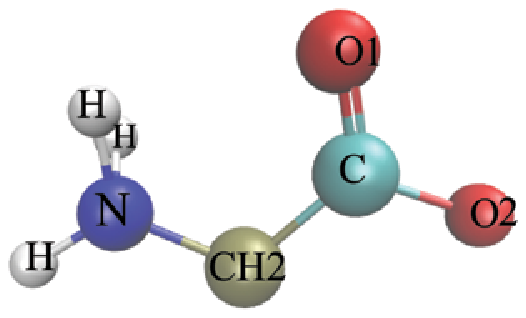}
\caption{Zwitterion glycine with CH2 as united atom centered at position of atom C. }\label{fig:gly}
\endminipage\hfill
\minipage{0.450\textwidth}
\centering
\includegraphics[scale=0.45]{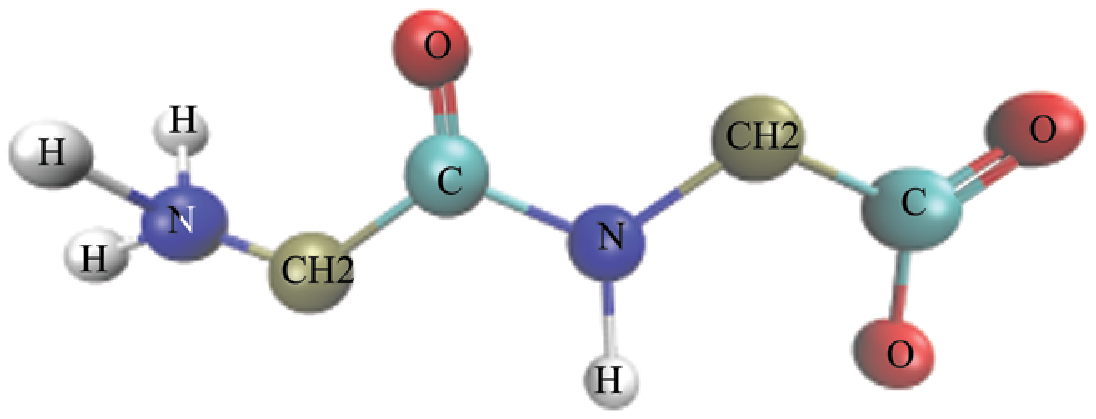}
\caption{Zwitterion diglycine with CH2 as united atom centered at position of atom C. }\label{fig:dgl}
\endminipage\hfill
\end{figure}

\begin{figure}[h]
\centering
\includegraphics[scale=0.5]{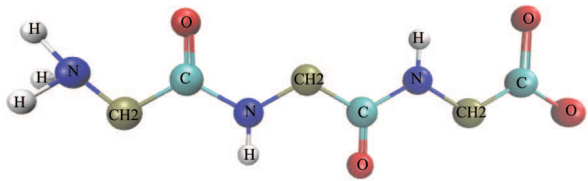}\hfill
\caption{Zwitterion triglycine with CH2 as united atom centered at position of atom C.}\label{fig:tgl}
\end{figure}

In the present work, zwitterion form of glycine, dyglycine and triglycine were taken and modeled in GROMOS53A6 force field platform \cite{forcefield1, forcefield2}. Specific bonds, and bond angles were taken in g96 format to configure the topology of the molecules. Different proper dihedrals were defined to prevent rotation around a bond and improper dihedrals were defined to confine four atoms in plane or tetrahedral configuration. We took CH$_2$ as united atom in which position of united atom is the position of heaviest atom, C in this case. Figure-\ref{fig:gly} shows the models of zwitterion glycine, figure-\ref{fig:dgl} shows  diglycine and figure-\ref{fig:tgl} shows triglycine molecules. SPC/E model of water \cite{SPCE} was taken as solvent. To estimate the diffusion coefficient of zwitterion glycine in water and check its dependency in temperature, two zwitterion glycine molecules were dissolved in 11,112 water molecules in system-I. It was simulated at four different temperatures: 293.2 K, 303.2 K, 313.2 K, and 333.2 K. In order to estimate variation of diffusion coefficient with increase in monomer units in peptide chain two zwitterion glycine, two zwitterion diglycine, and two zwitterion triglycine molecules were separately dissolved in 1,385 water molecules in system-II, system-III, and system-IV respectively and diffusion coefficients were estimated at temperature 298.2 K. All the simulations were carried out at pressure of one atmosphere. The number of molecules were chosen to match mole fraction of experimentally reported data \cite{ Umecky, longsworth2}.

\noindent During the simulation if the simulation box contains overlapped particles or particle with bad contact, molecular dynamics could explode and never bring system to equilibrium \cite{allen}. To ensure equilibration energy minimization was carried out for each of the systems with threshold energy of 50 kJ/mol. Steepest-descent method of energy minimization was used as its simple and sturdy, which simply takes a step in the direction of the negative gradient, without any consideration of the history built up in previous steps. The step size is adjusted such that the search is fast, but the motion is always downhill \cite {manual}. Particle Mesh Ewald (PME) was used for coulomb interaction. Cut off distance for both of the non bonded interactions- coulomb and LJ was 1.0 nm. Dynamical properties like diffusion, thermal conduction, etc depend strongly upon the temperature, and pressure of system \cite{sunildai}. To bring system under study at a condition that best mimics experimental environment each of the systems was equilibrated for 200 ns with time step of 0.02ps. Temperature of system was coupled to reference temperature using  velocity rescale and reference pressure was coupled to system using berendsen coupling. Density of the equlibrated system was tallied with experimentally reported values to  ensure proper equilibration and to be sure that the force field parameters used are suitable for the system under consideration. In no case, density of system after equilibration differed from experimental value by 1.5 percent. This suggests that the force field parameters used are well suited for the systems under consideration. After equilibration of system, production run was carried out to estimate diffusion coefficients. Structural of system after production run was explored using radial distribution function (RDF).

\section{RESULTS AND DISCUSSIONS}
\noindent We have calculated the self diffusion coefficients of glycine and water, and their binary diffusion coefficient at different temperatures using system-I. Diffusion coefficients of aqueous glycine, diglycine, and triglycine have been calculated at temperature 298.2 K. These diffusion coefficients have been calculated using Einstein method \cite{crank}, where self diffusion coefficient of any species in three dimensional isotropic medium is one sixth of slope of mean squared displacement (MSD) versus time plot as in equation (\ref{selfdiff}). Structure of systems have been studied using the radial distribution function (RDF), which gives preferred position of one particle around other particle.

\subsubsection*{Mean Squared Displacement (MSD)}
\begin{figure}[!htb]
\minipage{0.450\textwidth}
\includegraphics[scale=0.30]{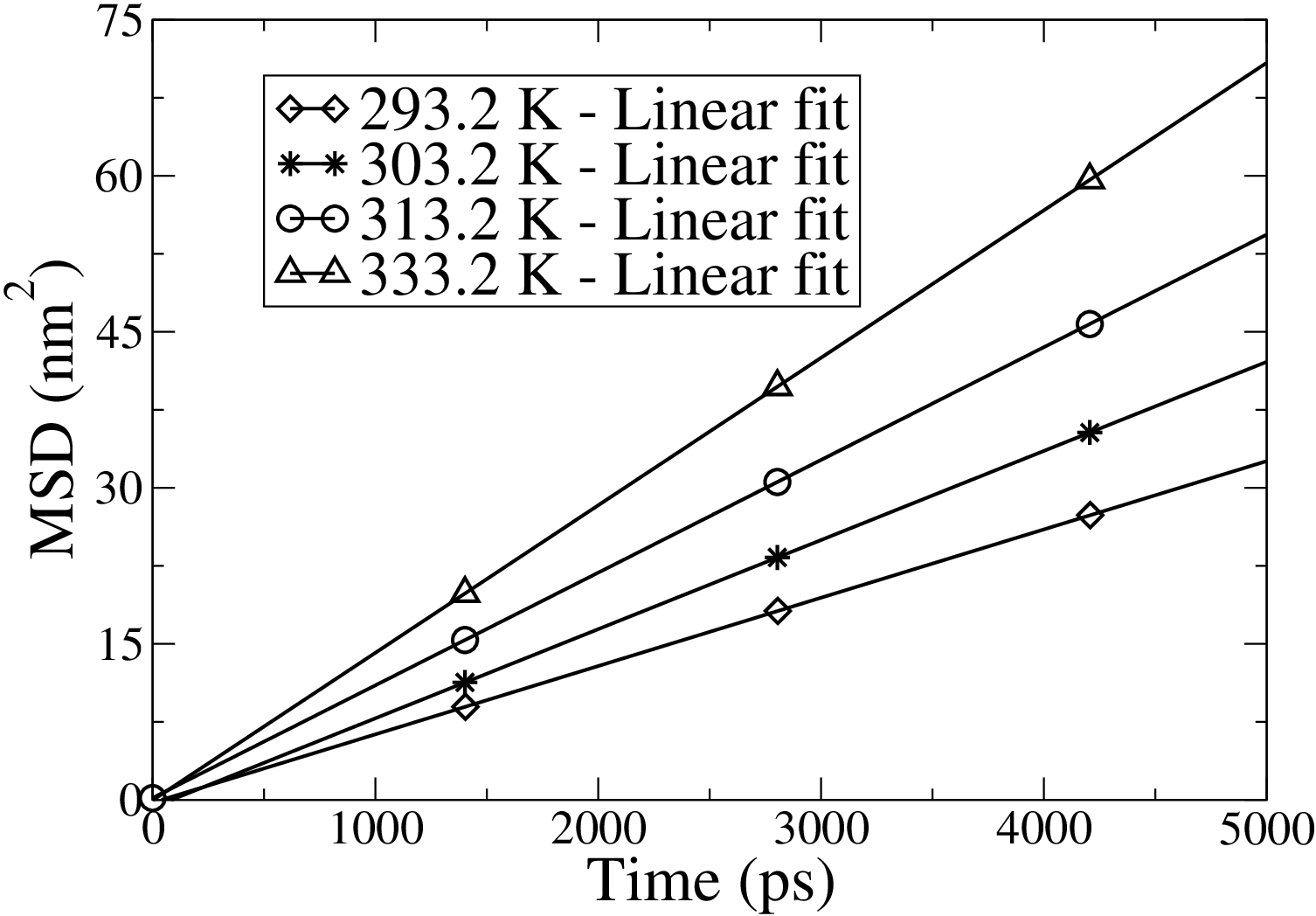}
\caption{MSD plot of glycine at four different temperatures. }\label{fig:msdgly}
\endminipage\hfill
\minipage{0.450\textwidth}
\includegraphics[scale=0.30]{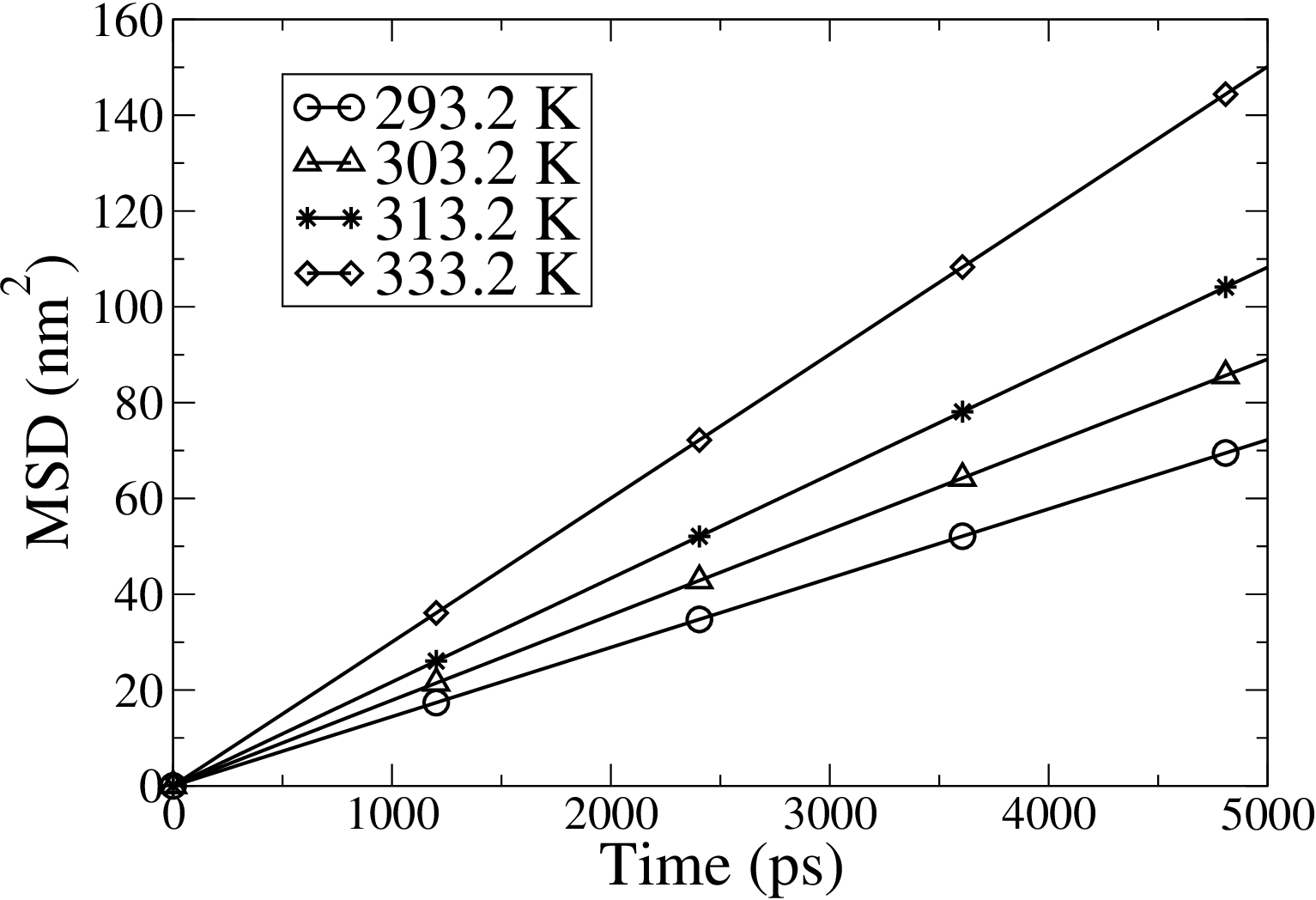}
\caption{MSD plot of water at four different temperatures.}\label{fig:msdwat}
\endminipage\hfill
\end{figure}

\begin{figure}[!htb]
\minipage{0.450\textwidth}
\includegraphics[scale=0.30]{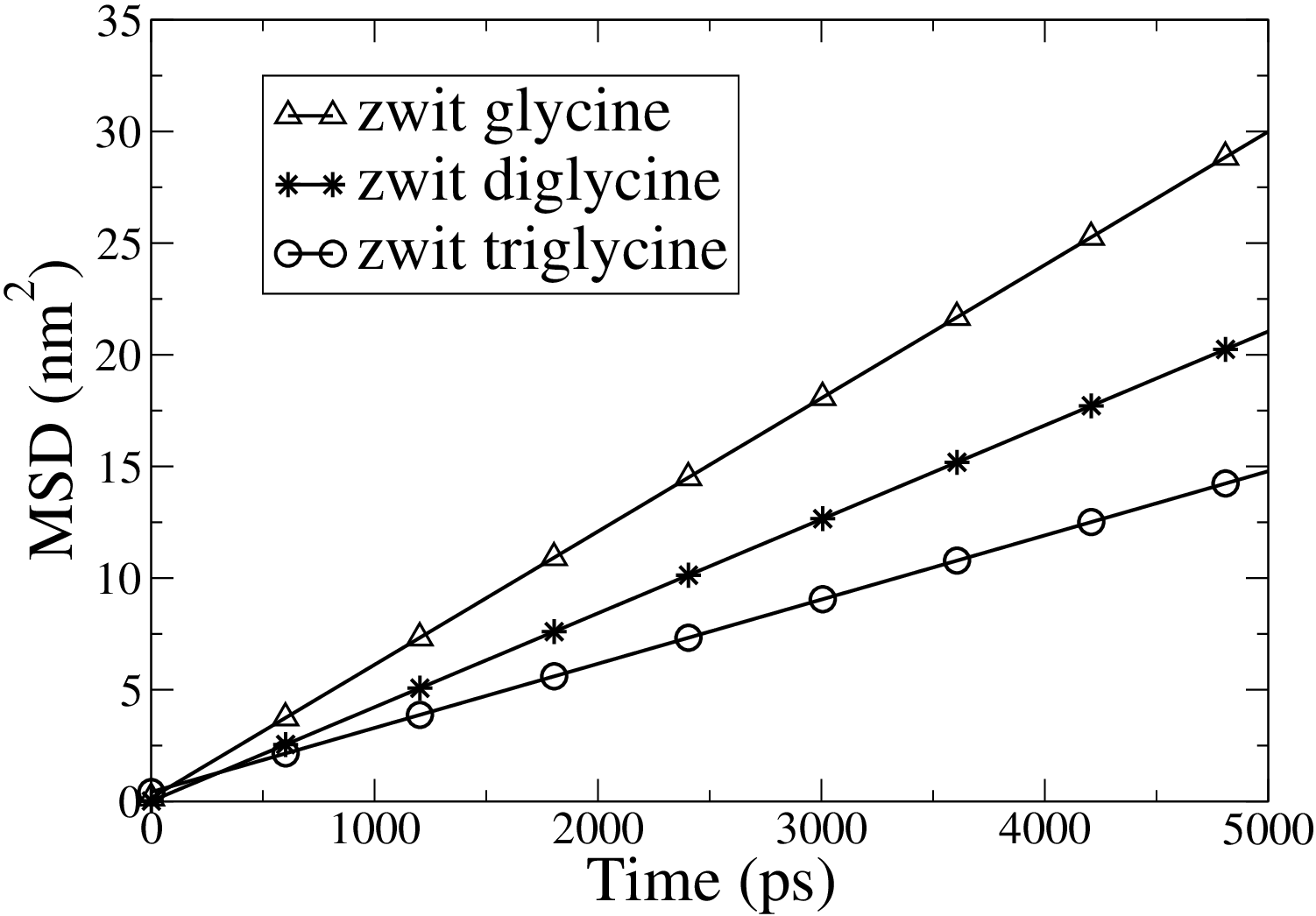}
\caption{MSD plot of glycine, diglycine and triglycine at temperature 298.2 K.}\label{fig:msdgdt}
\endminipage\hfill
\minipage{0.450\textwidth}
\includegraphics[scale=0.30]{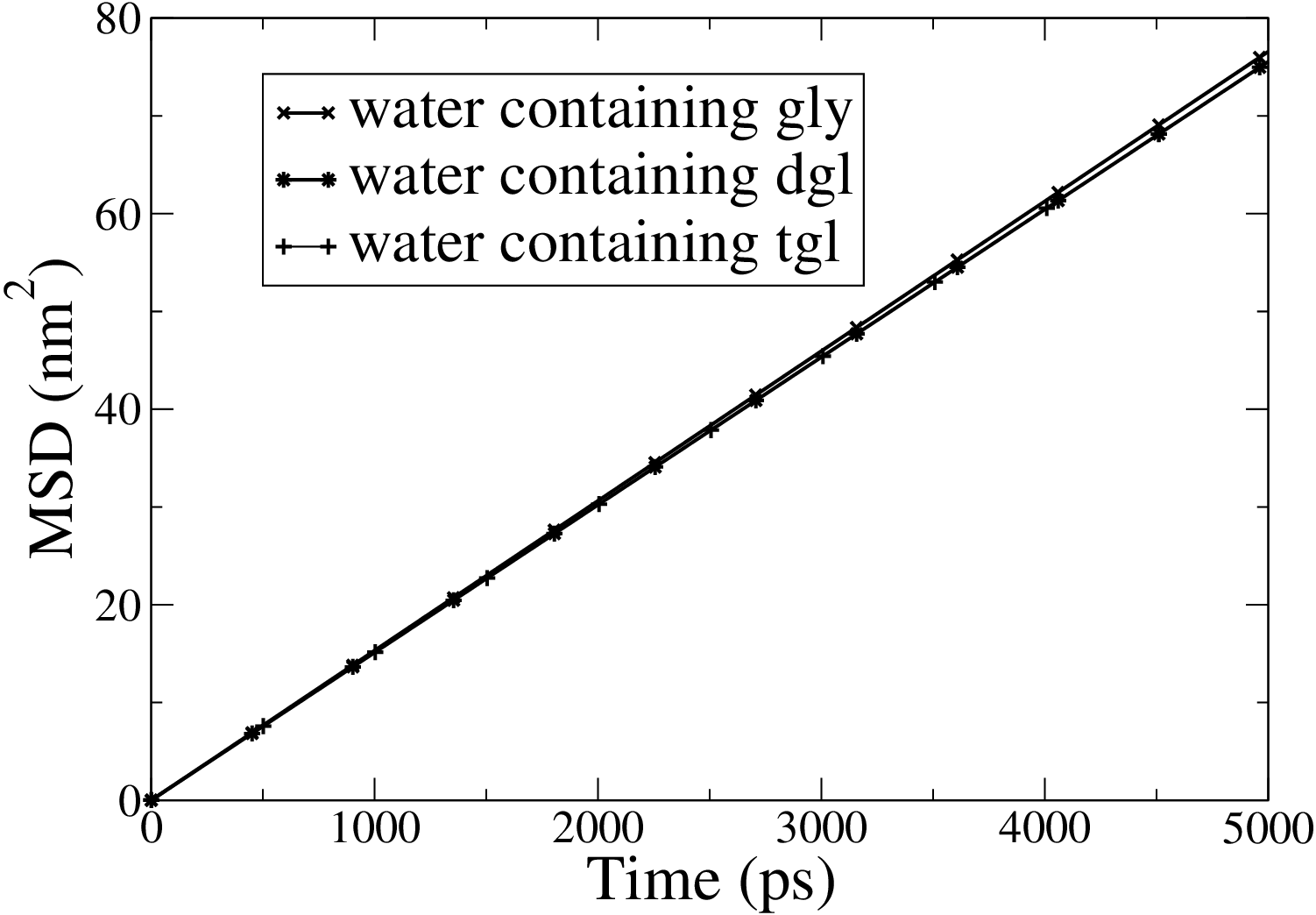}
\caption{MSD plot of water with different solute at temperature 298.2 K.}\label{fig:msdgdtwat}
\endminipage\hfill
\end{figure}

\noindent Mean squared displacement (MSD) as a function of time is used to calculate self diffusion coefficient in Einstein's method. Linear fit of MSD of different molecules at different temperatures are plotted as a function of time for 5 ns. Figure-\ref{fig:msdgly} is MSD plot of glycine at four different temperatures. The MSD is becoming steeper with increase in temperature. This shows that rate of diffusion is increasing with increase in temperature. Figure-\ref{fig:msdwat} is MSD plot of water at four different temperatures and shows similar behaviour of MSD plot of glycine in figure-\ref{fig:msdgly}. MSD plot of glycine, diglycine and triglycine at temperature 298.2 K is shown in figure-\ref{fig:msdgdt}. The MSD line is getting less steep with increase in molecule size indicating smaller rate of diffusion of larger molecule than that of smaller molecule at same temperature. MSD plot of water at 298.2 K containing three different solutes is shown in fiugre-\ref{fig:msdgdtwat}. The MSD curve is steeper when lighter solute molecules are present. It indicate that water diffuses faster at given temperature when smaller solute molecules are present. Figure-\ref{fig:msdgdt} and figure-\ref{fig:msdgdtwat} show the size effect of solute on MSD and diffusion. 

\subsubsection*{Self Diffusion Coefficient}
\noindent Table \ref{table:selfdiff} presents the estimated values of self diffusion coefficient of different molecules at different temperatures. The self diffusion coefficient of zwitterion glycine and water is increasing with increase in temperature. This is in accord with increased random velocity with increased temperature. Self diffusion coefficients of zwitterion glycine, zwitterion diglycine and zwitterion triglycine each dissolved separately in 1,385 water molecules at temperature 298.2 K and pressure 1 bar is presented in the same table. The estimated self diffusion coefficient of these molecules have decreased with increase in monomer unit in chain. It could be attributed to size of the molecules. At given temperature molecule of larger molecular weight attains smaller velocity than lighter molecule. Therefore, in the same environment and physical conditions, large molecules diffuses slowly than small molecule. Chain length and hence molecular weight of molecule increases from zwitterion glycine, zwitterion diglycine to zwitterion triglycine and the rate of diffusion decreases. At temperature 298.2 K, self diffusion coefficient of glycine appears less than that at 293.2 K. It could be because of higher concentration of glycine at 298.2 K compared to that at 293.2 K. 

\begin{table}[h]
\centering
\caption{Estimated values of self diffusion coefficient of diferent molecules at one atm pressure and different temperatures. }
\label{table:selfdiff}
\begin{tabular}{|c|c|c|c|c|}
\hline
\multirow{2}{*}{Temperature (K)} & \multirow{2}{*}{Solute} &\multirow{2}{*}{ $D_{self}^{solute}(10^{-9}m^2.s^{-1})$} & \multicolumn{2}{|c|}{Water$(10^{-9}m^2.s^{-1})$}\\\cline{4-5}
															& & &  $D_{self}^{est}$ & $D_{self}^{exp}$\\
\hline
293.2 & Glycine &1.09& 2.48 & 2.03 \cite{waterexpe}\\
\hline
\multirow{3}{*}{298.2} & Glycine 		& 0.99& 2.55 &2.30 \cite{waterexpe}  \\\cline{2-5}
					   & Diglycine      & 0.70& 2.52 &2.30 \cite{waterexpe}  \\\cline{2-5}
					   & Triglycine     & 0.48& 2.51 &2.30 \cite{waterexpe}  \\\cline{2-5}			   					
\hline 
303.2 & Glycine & 1.43	&2.97 	&2.59 \cite{waterexpe}  \\
\hline
313.2 & Glycine & 1.81   &3.61 	& 3.24 \cite{waterexp} \\
\hline
333.2 & Glycine & 2.36 &5.00	& 4.77 \cite{waterexp} \\
\hline
\end{tabular}
\end{table}

\noindent The value of self diffusion coefficient  of water has increased with increase in temperature. Estimated values are slightly greater than experimental values reported in literatures. It has been shown by Paudyal et al. \cite{Ipaudyal} that diffusion coefficient of water estimated using GROMACS slightly increases with increase in size of simulation and they have shown that estimated values best tally with experimental values at small simulation size. All the systems in the present work contained large number of water molecules. That could be the reason why estimated values here are slightly higher than experimental values. Further, the self diffusion coefficient could have been affected by presence of solute. The self diffusion coefficient of water at 298.2 K is slightly smaller in the presence of larger molecules because of increased hindrance in motion of water molecules with increased solute size.

\subsubsection*{Binary Diffusion Coefficient}
\noindent Binary diffusion coefficient has been calculated for zeitterion glycine-water mixture at four different temperatures, and mixture of zwitterion glycine - water, zwitterion diglycine - water and zwitterion triglycine - water at temperature 298.2 K using Darken's relation (\ref{darken}). Binary diffusion coefficients of different pairs of molecules at different temperatures are presented in table-\ref{table:binarydiff} and compared with experimental values reported in literature \cite{Umecky, longsworth2}. $D_{binary}^{est}$ is binary diffusion coefficient estimated in present work and $D_{binary}^{exp}$ is experimental value of binary diffusion coefficient reported in literatures. With the increase in temperature, thermal agitation of molecules increases which boosts the diffusion. This means diffusion coefficient should be greater at higher temperatures.\\

\begin{table}[!h]
\centering
\caption{Binary diffusion coefficient of different paris of molecules at different temperatures.}
\label{table:binarydiff}
\begin{tabular}{|c|c|c|c|c|c|}
\hline
\multirow{2}{*}{Solvent} &\multirow{2}{*}{Solute}  & \multirow{2}{*}{Temperature (K)}&\multicolumn{2}{|c|}{ Binary diffusion coeff. $m^2/s$}& $\%$ Error\\\cline{4-5}
& & & $D_{binary}^{est} (10^{-9}m^2.s^{-1})$& $D_{binary}^{exp}(10^{-9}m^2.s^{-1})$& \\
\hline
\multirow{7}{*}{Water} & Glycine 	&  	293.2 &1.10 &0.94 \cite{Umecky}& 17.02 \\\cline{2-6}
                       & Glycine 	&  	298.2 &0.99 &1.06 \cite{longsworth2}& 6.60 \\\cline{2-6}
                       & Diglycine 	&  	298.2 & 0.70  & 0.79 \cite{longsworth2}& 11.39 \\\cline{2-6}
                       & Triglycine & 	298.2 & 0.48& 0.67 \cite{longsworth2}& 28.36 \\\cline{2-6}
                       & Glycine 	&	303.2 & 1.43 & 1.22 \cite{Umecky}& 17.21 \\\cline{2-6}
                       & Glycine	& 	313.2 & 1.81 & 1.50 \cite{Umecky}& 20.67 \\\cline{2-6}
                       & Glycine 	& 	333.2 & 2.36   & 2.22 \cite{Umecky} & 6.31 \\\cline{2-6}
 
\hline
\end{tabular}
\end{table} 
 
\noindent Binary diffusion coefficient of glycine in water is increasing with increase in temperature, just like self diffusion coefficient. Binary diffusion coefficient of glycine, diglycine and triglycine in water is also follwoing decreasing nature of self diffusion coefficient with increase in molecular weight.  Mass of molecules increases in progressing from glycine to diglycine and then to triglycine and it is seen that the value of binary diffusion coefficient decreases sharply with increase in number of monomers in peptide chain. The standard error in the estimation of self as well as binary diffusion coefficients are very small and insignificant compared to the estimated values where as symmetric round off errors in each estimation of diffusion coefficient is of the order $10^{-11}$. The value of D$_{binary}^{ext}$ for triglycine has deviated from experimentally reported value by about 28 percent. It is possible that the united atom modeling and corresponding force field parameters we used might not be adequate for large molecule like triglycine. Moreover, the data in the literature \cite{longsworth2} have been reported using Rayleigh interference method. We suggest to use more advanced and reliable experimental methods like NMR to measure diffusion coefficient.\\ 

\subsubsection*{Effect of Temperature on Diffusion}
\begin{figure}[h]
\minipage{0.45\textwidth}
\includegraphics[scale=0.30]{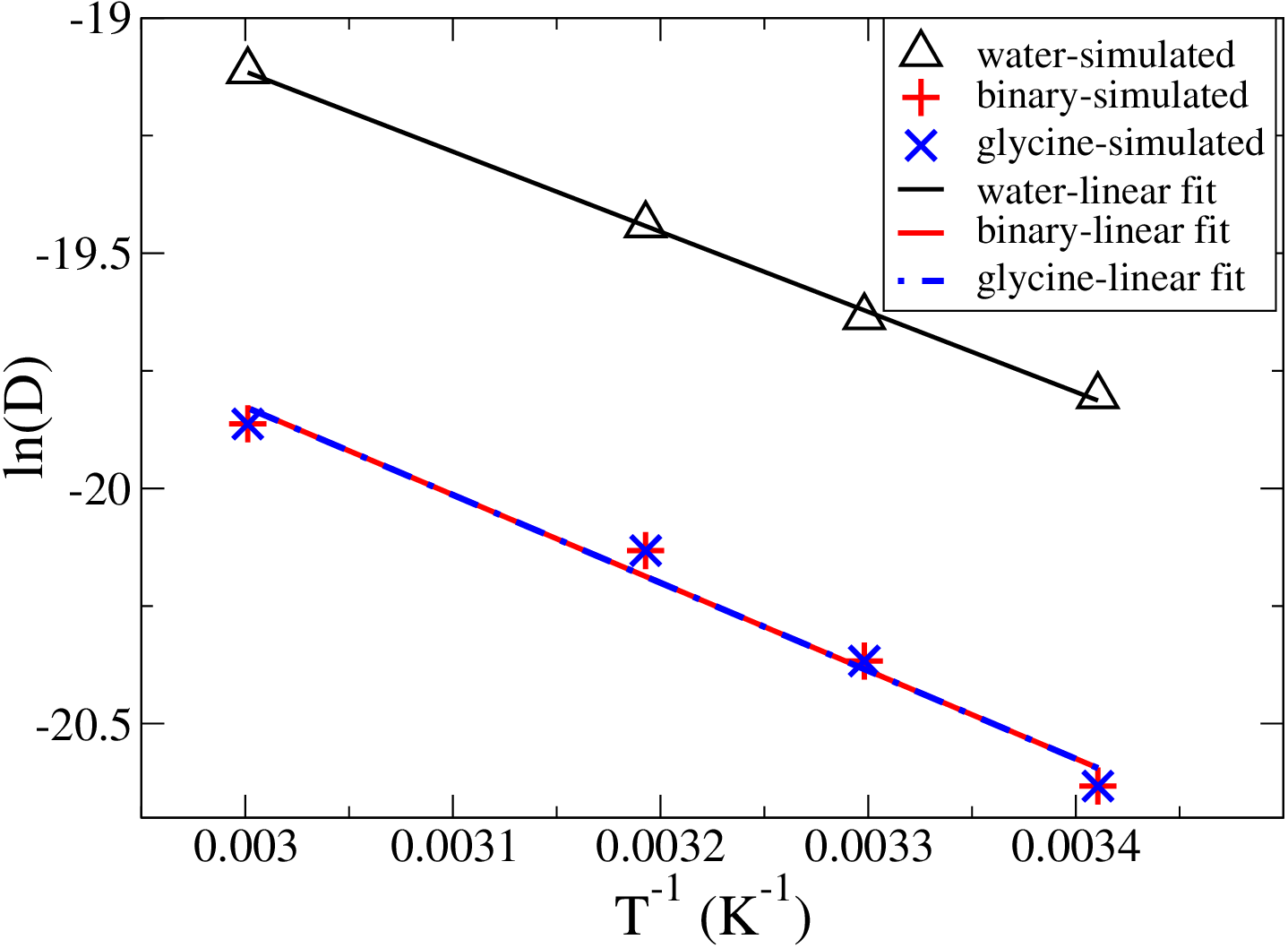}
\caption{Arrhenius diagram of diffusion coefficient of water, zwitterion glycine, and their binary mixture.}\label{fig:arrheniusall}
\endminipage\hfill
\minipage{0.450\textwidth}
\includegraphics[scale=0.3]{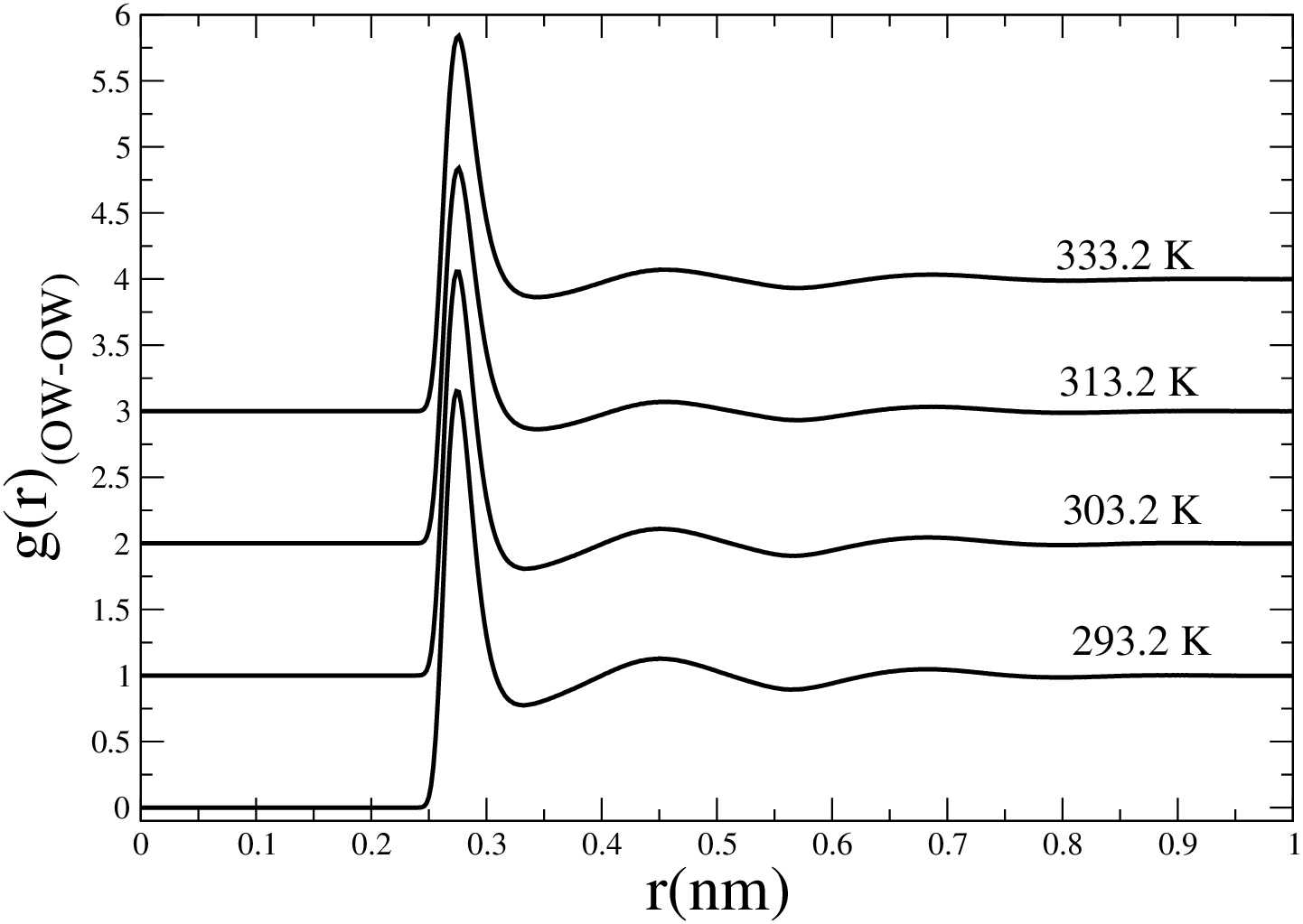}
\caption{Radial distribution function of oxygen of water (OW) in reference of oxygen of water (OW).}\label{rdf:owow}
\endminipage\hfill
\end{figure}

Diffusion is transport of mass due to concentration and thermal gradient. It strongly depends upon temperature \cite{udhal}. We have checked the temperature dependency of diffusion coefficient of glycine, water and their binary mixture using Arrhenius formula \cite{crank}:
\begin{equation}
\label{arrhenius}
 D =  D_o \exp( -\frac{E_a}{N_A k_B T})
\end{equation}

\noindent where D$_o$ denotes the pre-exponential factor, also called the frequency factor, E$_a$ is the activation energy for diffusion, T is the absolute temperature, N$_A$ is Avogadro number whose value is 6.022x10$^{23}$ per mol, and k$_B$ is the Boltzmann constant whose value is 1.38x10$^{-23}$ J.K$^{-1}$. The activation energy of diffusion process corresponds to the slope of Arrhenius plot, which is plot between ln(D) and reciprocal of absolute temperature. Figure \ref{fig:arrheniusall} shows the Arrhenius diagram of self-diffusion coefficient of zwitterion glycine, self-diffusion coefficient of water, and their binary-diffusion coefficient. In the plot, points obtained from simulation have aligned around straight line. Hence, it could be said that diffusion coefficient of these molecules at different temperatures follows Arrhenius behaviour. As the self diffusion coefficient of glycine and binary diffusion coefficient of aqueous glycine are nearly equal, Arrhenius plot have overlapped. This is because of very small mole fraction of glycine. The pre-exponential factors and  activation energies are presented in table-\ref{table:arrhenius}.\\


\begin{table}[H]
\centering
\caption{Pre-exponential factor and activation energy of diffusion.}
\label{table:arrhenius}
\begin{tabular}{|c|c|c|}
\hline
Molecule& Pre-exponent factor (D$_o$ $m^2.s^{-1}$ )&Activation energy (E$_a$)(kJ.mol$^{-1}$) \\
\hline
	Water &  8.23 x $10^{-7}$  &14135.48 \\
\hline
Zwit. glycine & 6.67 x $10^{-7}$ &15517.51 \\
\hline
Binary mixture- simulated &6.74 x $10^{-7}$ &15517.18 \\
\hline
\end{tabular}
\end{table}

\subsubsection*{Radial Distribution Function (RDF)}
\noindent Radial distribution function (RDF) is used to study the pair correlation and structure of system. It gives the preferred position of one particle with respect to other particle. For isotropic system, it is only the function of distance between particles \cite{mcquarrie}.\\

\noindent Figure-\ref{rdf:owow} is the RDF of oxygen atom of water (OW) in reference to  oxygen atom of other water molecule (OW). van der Waals radius of oxygen atom is 2$^{ 1/6} \sigma$ = 0.355 nm \cite{DipendraPaper}. This means, if these atoms were to interact only under LJ potential, they wouldn't go closer to each other than van der Waals radius. In the plot first peak positions are 0.274 nm, 0.275 nm, 0.276 nm, and 0.276 nm and corresponding peak values are 3.151, 3.053, 2.982, and 2.835 at temperatures 293.2 K, 303.2 K, 313.2 K, and 333.3 K respectively. The hydrogen and oxygen atom in SPC/E model of water has partial positive and negative charges. Thus, the coulomb interaction is responsible for first peak position being smaller than van der Waals radius. Value of r in this plot gives the preferred distance of oxygen atom in a water molecule around other water molecules and value of g(r) gives relative the probability of finding OW. Only three peaks are observable and beyond that value of g(r) is unity which means there is no correlation between oxygen in water molecules beyond the position of third peak.\\

\begin{figure}[h]
\minipage{0.450\textwidth}
  \includegraphics[scale=0.3]{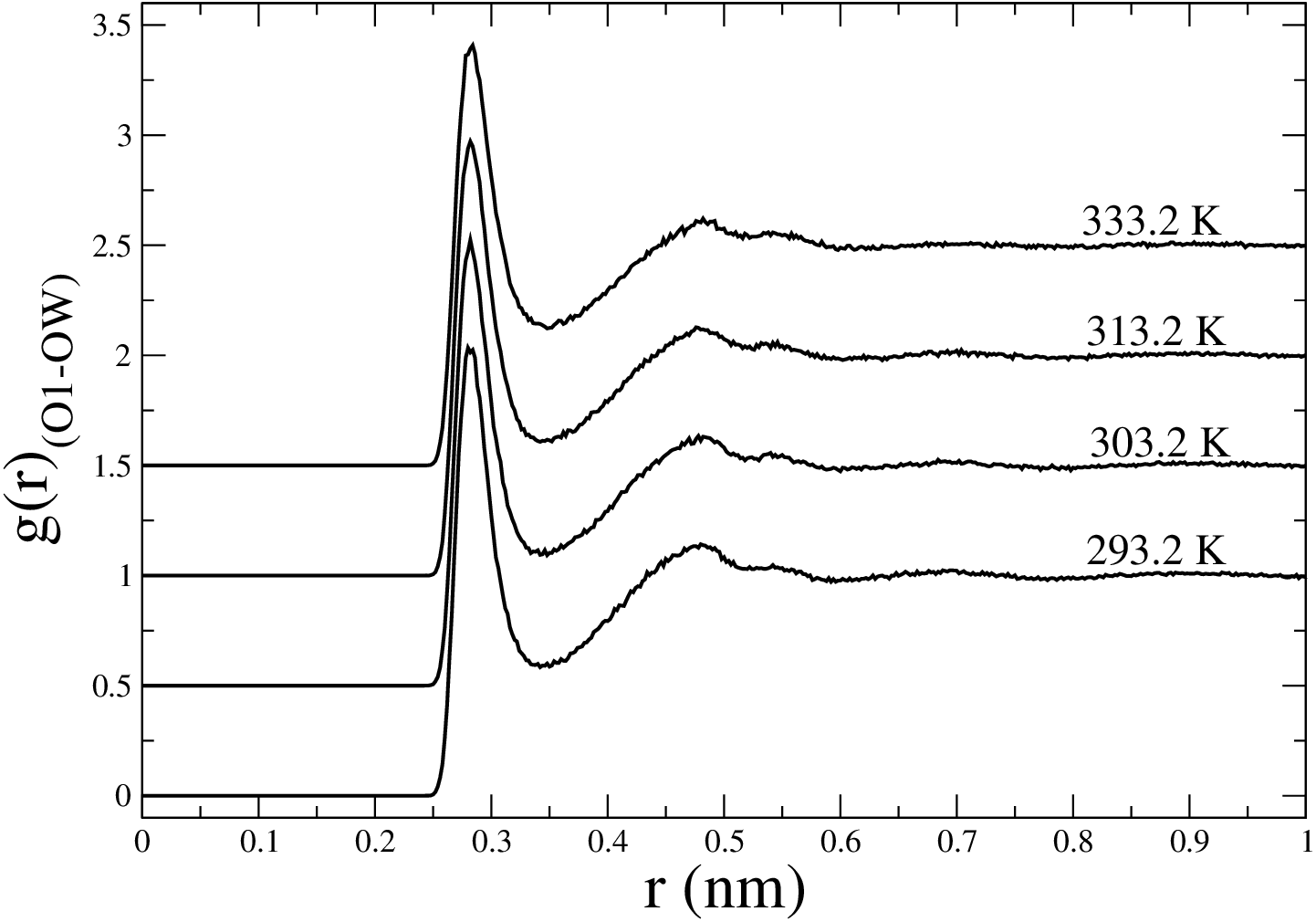}
\caption{Radial distribution function of oxygen (OW) in water molecule in reference of oxygen (O1) in zwitterion glycine.}\label{rdf:o1ow}
\endminipage\hfill
\minipage{0.450\textwidth}
  \includegraphics[scale=0.3]{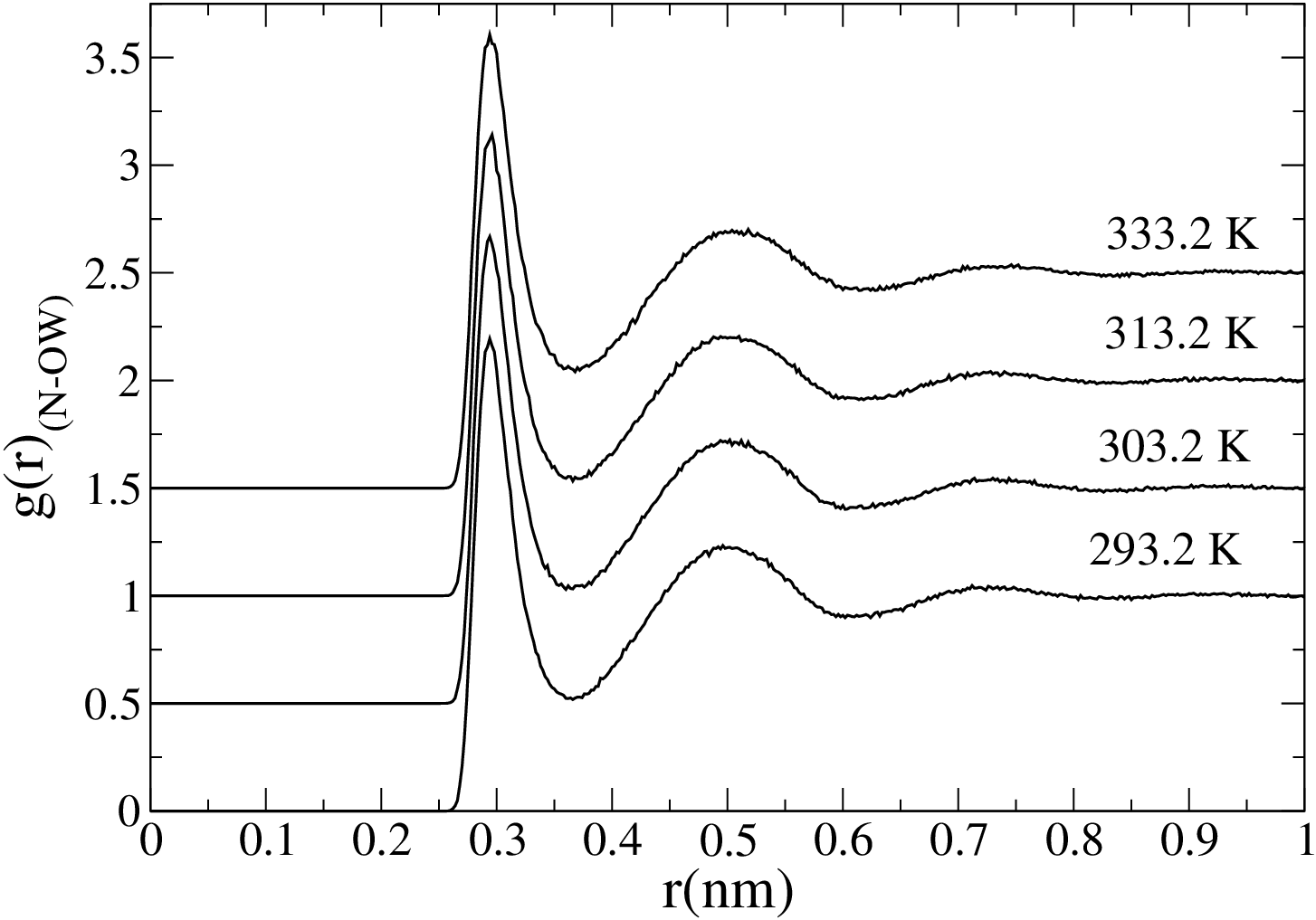}
\caption{Radial distribution function of oxygen of water (OW) in reference of nitrogen (N) in zwitterion glycine.}\label{rdf:now}
\endminipage\hfill
\end{figure}

\noindent Figure-\ref{rdf:o1ow} is the RDF of oxygen (OW) in water molecule in reference of oxygen (O1) in glycine. It gives average distribution of OW and hence the water molecules around O1 in zwitterion glycine. The first peak positions of OW around O1 at temperatures 293.2 K, 303.2 K, 313.2 K, 333.2 K are 0.280 nm, 0.282 nm, 0.282 nm, and 0.284 nm while the corresponding peak values are 2.034, 2.024, 1.971, and 1.907 respectively. Figure-\ref{rdf:now} is the RDF of OW in reference of nitrogen (N) in glycine. It gives the distribution of water molecules around $NH3^+$ terminal of zwitterion glycine. The first peak positions of OW around N at temperatures 293.2 K, 303.2 K, 313.2 K, 333.2 K are 0.294 nm, 0.294 nm, 0.296 nm, and 0.298 nm while the corresponding peak values are 2.191, 2.161, 2.128, and 2.062respectively. \\

\noindent In all three RDF plots, with increase in temperature, the first peak positions have moved farther while  hight of first peak have decreased and their width have increased. These phenomenon indicate the increased random motion with temperature. Further, wide RDF at higher temperatures means more space in between the molecules. This allows the molecules to move more freely, resulting increase in diffusion coefficient. Therefore, the increase in width of RDF means increase in diffusion. The existence of many peaks in RDF, their position, height and width could not be explained just by interaction between pair of species under calculation but interaction between and of these species with other species in the surrounding.

\section{Conclusions and Concluding Remarks}
\noindent We carried out realistic classical molecular dynamics simulation of glycine, diglycine, and triglycine in water where concentration of solutes were same as that reported in experiments whose results were compared with present work. The estimated values were in good agreement with experimental datas. The solutions used in simulation were very dilute and  binary diffusion coefficients were nearly equal to binary diffusion coefficient of molecule of extremely small mole fraction. Temperature dependency of diffusion coefficient of glycine, water and their binary mixture was tested. RDF revealed distribution of different atoms in isotropic medium and effect of temperature on diffusion. 

\section*{Acknowledgement}
YPK acknowledges the Master Thesis Grants from University Grants Commission, Nepal.

\end{document}